\documentclass[doublecol]{epl2} 

\usepackage{epsfig}

\usepackage{amssymb}
\usepackage{amsmath}
\usepackage{amsfonts}
\usepackage{bbold}
\usepackage{bm}
\usepackage{graphicx}

\newcommand{\nn}{\nonumber}
\newcommand{\ud}{{\textrm{d}}}
\newcommand{\bk}{{\bf k}}
\newcommand{\br}{{\bf r}}
\newcommand{\bg}{{\bf g}}
\newcommand{\bI}{{\bf I}}
\newcommand{\bA}{{\bf A}}

\newcommand{\llangle}{\langle\kern-.25em\langle}
\newcommand{\rrangle}{\rangle\kern-.25em\rangle}

\title{Replica symmetry breaking in the Bose glass}
\shorttitle{Replica symmetry breaking in the Bose glass} 

\author{S.~J. Thomson\inst{1,2} \and F. Kr\"uger\inst{1,2,3}}
\shortauthor{S.~J. Thomson and  F. Kr\"uger}

\institute{                    
  \inst{1} SUPA, School of Physics and Astronomy, University of St~Andrews, North Haugh, St~Andrews, KY16 9SS, United Kingdom \\
  \inst{2} ISIS Facility, Rutherford Appleton Laboratory, Chilton, Didcot, Oxfordshire OX11 0QX, United Kingdom\\
  \inst{3} London Centre for Nanotechnology, University College London, Gordon St., London, WC1H 0AH, United Kingdom 
}


\abstract{We investigate the nature of the Bose glass phase of the disordered Bose-Hubbard model in $d>2$ and demonstrate the existence 
of a glass-like replica symmetry breaking (RSB) order parameter in terms of particle number fluctuations. Starting from a strong-coupling expansion around the atomic limit, 
we study the instability of the Mott insulator towards the formation of a Bose glass. We add some infinitesimal RSB, following the Parisi hierarchical approach
in the most general form, and observe its flow under the momentum-shell renormalization group scheme. We find a new fixed point with one-step RSB,
corresponding to the transition between the Mott insulator and a Bose glass phase with hitherto unseen RSB. The susceptibility associated to infinitesimal RSB 
perturbation in the Mott insulator is found to diverge at 
the transition with an exponent of $\gamma=1/d$. Our findings are consistent with the expectation of glassy 
behavior and the established breakdown of self-averaging.  We discuss the possibility of measuring the glass-like order parameter in optical lattice experiments as well as 
in certain spin systems that are in the same universality class as the Bose-Hubbard model.}

\begin{document}

\maketitle

\section{Introduction}
Bosons in optical lattices offer an exciting opportunity for the study of disorder in strongly-interacting systems. In the clean Bose-Hubbard (BH) 
model two phases compete. The Mott insulator (MI) wins out when the on-site repulsion is sufficiently large, whereas the superfluid (SF) forms when the 
kinetic energy of the particles dominates. In the disordered BH model a third phase enters the fray: the gapless and insulating Bose glass (BG) 
which has been predicted  \cite{Fisher+89}  to exist between the MI and SF phases. Despite intense analytical \cite{Giamarchi+88,Singh+92, Mukhopadhyay+96,Freericks+96,Svistunov96,Herbut97,Wu+08,Kruger+09,Weichman+08,Bissbort+09,Iyer+12,Niederle+13} and 
numerical \cite{Scalettar+91, Krauth+91,Pai+96,Sen+01,Lee+01,Prokof'ev+04,Kisker+97} study, the 
question of whether a direct MI-SF transition is possible in the presence of disorder has been controversial. Only recently has it been proven 
\cite{Pollet+09,Gurarie+09} that the BG always intervenes between the MI and the SF, regardless of how weak the disorder. 

The nature of the BG phase itself is still not very well understood. With recent advances in optical lattice technology \cite{Sherson+10,Bakr+10,Weitenberg+11,Endres+11} 
and the ability to introduce disorder in carefully controlled ways \cite{Billy+08,White+09,Pasienski+10,Kondov+11,Pasienski+08,Bruce+11,Gaunt+12}, the time is 
now ripe for a detailed theoretical and experimental investigation into the properties of this phase. It has been argued by various 
authors \cite{Fisher+89,Kisker+97,Pollet+09,Niederle+13} that the BG is a Griffiths phase \cite{Vojta10}, with the behavior dominated by the existence 
of rare SF regions. As pointed out recently \cite{Kruger+11,Hegg+13}, the existence of rare SF regions is intimately related to a breakdown of self-averaging. 
Although the lack of ergodicity and self-averaging are common features of conventional spin glasses \cite{Binder+86,Fischer+91} and despite the fact that the BG 
is dubbed a `glass', an Edwards-Anderson-like glassy order parameter has not yet been shown to exist in this phase. 

In this Letter we establish such an order parameter within the replica framework. Performing a strong-coupling expansion and a 
momentum-shell renormalization group (RG) analysis, we investigate the instability of the MI towards the formation of the BG, allowing for Parisi RSB \cite{Parisi79}. 
We show that unlike in typical spin-glass 
systems, RSB in the BG phase only occurs on the quartic level, corresponding to glassy behavior in the particle density fluctuations rather than in the 
SF order parameter.  Our results are consistent with previous findings showing RSB in random-mass disordered ferromagnets \cite{Dotsenko+95a,Dotsenko+95b} and in the Anderson glass \cite{Giamarchi+96,Giamarchi+01}, the fermionic analog of the BG. Note that the incompressible
Mott glass exhibits RSB as well \cite{Giamarchi+01} suggesting that there exists no direct relation between the finite compressibility of the disordered insulator and the breaking of replica symmetry.

\section{Strong coupling expansion and replica theory}
Our starting point is the disordered BH model in its simplest form which describes bosons at chemical potential $\mu$ that tunnel with amplitudes 
$t$ between neighboring sites of a $d$-dimensional hyper-cubic lattice with random site energies $\epsilon_i$ and interact through a local Hubbard repulsion $U$. 
The Hamiltonian is given by
\begin{equation}
\hat{\mathcal{H}}=-t \sum_{\langle ij \rangle}(\hat{b}^{\dagger}_{i}\hat{b}_{j}+\hat{b}^{\dagger}_{j}\hat{b}_{i})-\sum_{i}\mu_i\hat{n}_{i}+\frac{U}{2}\sum_{i}\hat{n}_{i}(\hat{n}_{i}-1),
\end{equation}
where $\hat{b}^{\dagger}_{i}$, $\hat{b}_{i}$ are bosonic creation and annihilation operators, $\hat{n}_{i}=\hat{b}^{\dagger}_{i}\hat{b}_{i}$ the number 
operator, and $\mu_i=\mu-\epsilon_i$. The random site energies are uncorrelated 
and follow a box distribution $p(\epsilon_i)=\theta(\Delta-|\epsilon_i|)/2\Delta$ ($\theta$ is the Heaviside step function). For sufficiently bounded 
disorder, $\Delta/U<1/2$, the model exhibits MI phases.

The strong-coupling expansion around the atomic limit is based on a Hubbard-Stratonovich transformation of the hopping term 
\cite{Sengupta+05}. In the absence
of disorder, this leads to the imaginary time long-wavelength action,
\begin{equation}
\mathcal{S} = \int_{\bk,\omega}\left[k^2 +F(\omega)  \right] |\psi(\bk,\omega)|^2 + h \int_{\br,\tau}|\psi(\br,\tau)|^4,
\end{equation}  
with $\psi(\br,\tau)$ the complex SF order parameter which is a function of position $\br$ and imaginary time $\tau$. For convenience, we have Fourier transformed the quadratic 
action to the momentum and frequency domain, $(\bk,\omega)$. The function $F(\omega)=R+f(\omega)$ with $R=F(0)$  is related to the bosonic Green
function
\begin{equation}
\mathcal{G}_0(\omega)=-\frac{m+1}{m-x+i\omega}-\frac{m}{1-m+x-i\omega}
\end{equation} 
of the local on-site Hamiltonian, $F(\omega) = 1+y \mathcal{G}_0(\omega)$. 
Here $m$ denotes the number of bosons per site, $x=\mu/U$ and $y=2dt/U$ are the chemical potential and the hopping amplitude in dimensionless units, 
respectively. We have further rescaled lengths such that the momentum cut-off is given by $|\bk|\le\Lambda=1$. Except for isolated values of the chemical 
potential where the system is particle-hole symmetric, at $T=0$ the vertex $h$ is irrelevant in dimensions $d>2$. The transitions between the MI lobes and 
the SF are therefore of the mean-field type and obtained by $R=0$ for different integer values of $m$. 

In the presence of on-site disorder, the Hubbard-Stratonovich transformation can be performed in exactly the same way. All coefficients in the resulting 
strong-coupling action are given by the same expressions as in the clean case but with the chemical potential shifted by the 
disorder potential on each lattice site, e.g. the induced random-mass disorder is given by  
\begin{equation}
R_i = 1-y\left(\frac{m+1}{m-x+\epsilon_i} +\frac{m}{1-m+x-\epsilon_i} \right).
\label{eq.random_mass}
\end{equation}

Note that this expression is ill defined if $p(\epsilon_i)$ is not sufficiently bounded. In order to average the free energy over the quenched disorder, we use the replica
trick, $\overline{F}=-T \overline{\ln \mathcal{Z}} = \lim_{n \rightarrow 0} (\overline{\mathcal{Z}^n}-1)/n$. After taking $n$ copies of the system and performing the 
average over disorder we obtain the effective continuum action
\begin{eqnarray}
\mathcal{S} & = & \sum_{\alpha=1}^n\int_{\bk,\omega}\left[k^2 + \overline{R}+f(\omega)  \right] |\psi_\alpha(\bk,\omega)|^2+\ldots\nn\\
& & -\frac12 \sum_{\alpha\beta} g_{\alpha\beta}\int_{\br,\tau,\tau'} |\psi_\alpha(\br,\tau)|^2 |\psi_\beta(\br,\tau')|^2.
\label{eq.replica_action}
\end{eqnarray}

The replica diagonal contributions represent $n$  copies of the system with disorder averaged coupling constants while the disorder vertex $g_{\alpha\beta}$
couples different replicas. Because of the two independent integrations over imaginary time this term is relevant in dimensions $d<4$. Note that the replica disorder 
average and the cumulant expansion underlying Eq.~(\ref{eq.replica_action}) do not break the symmetry between replicas. All elements in the matrix $g_{\alpha\beta}$
are therefore identical and equal to the {\it variance} $g=\overline{\left(R-\overline{R}   \right)^2}$ of the random mass distribution (\ref{eq.random_mass}).

We point out that the disorder averaged replica action (\ref{eq.replica_action}) depends only on the first two moments of the random-mass distribution $R_i$.
Higher moments enter in the coefficients of terms beyond quartic order which are irrelevant under the RG. However, the first and second moments of $R_i$
depend on the precise form of the microscopic disorder distribution $\epsilon_i$, in other words on \emph{all} the moments of $\epsilon_i$. The existence of a Mott insulating
state, which is essential for the above expansion, requires $\epsilon_i$ to be bounded.

\section{Parisi replica symmetry breaking} 
\begin{figure}[t!]
\begin{center}
\includegraphics[width= \linewidth]{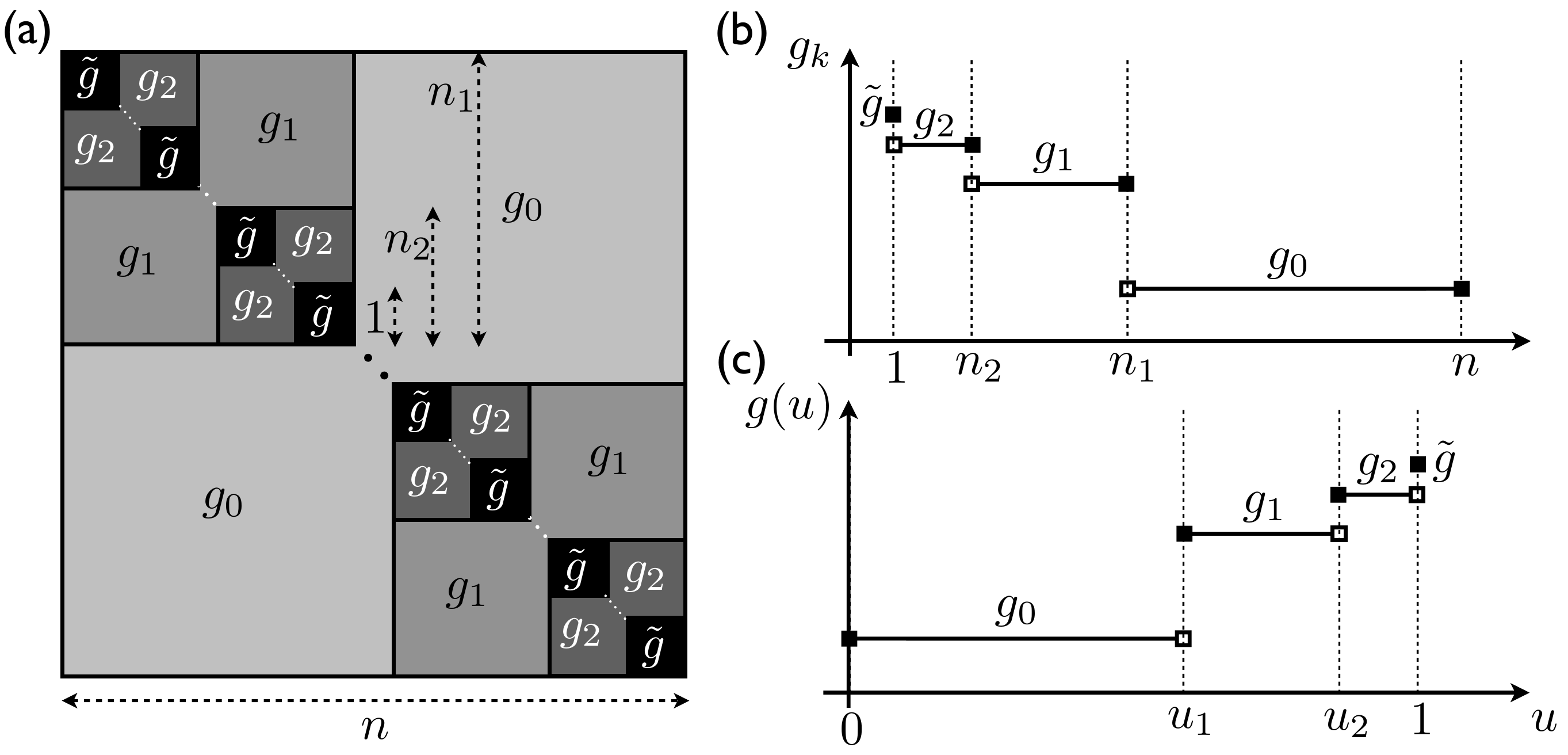}
\caption{(a) Parisi matrix $\bg$ with two-step RSB. (b) Matrix elements along the first row which usually decrease with distance from the diagonal. (c) In the replica 
limit $n\to 0$, the off-diagonal elements are described by a 2-step function $g(u)$ on the interval $u\in[0,1)$.}
\label{fig.1}
\end{center}
\end{figure}

In order to study the instability of the system towards RSB we allow $\bg=(g_{\alpha\beta})$ to be a general matrix of the Parisi type. 
Such matrices are constructed by successively introducing smaller and smaller block matrices along the diagonal. To give a formal definition of a Parisi matrix $\bg$
with $k$-step RSB, we introduce a sequence of integers $1=:n_{k+1}<n_k<\ldots <n_2<n_1<n_0:=n$ with $n_i/n_{i+1}$ integer and auxiliary $n\times n$ matrices 
$\bA_i$ which have $n_i\times n_i$ block matrices of ones along the diagonal and zeros outside these blocks. The Parisi matrix is then given by 
$\bg=\tilde{g}\bI+\sum_{i=0}^k g_i(\bA_{i+1}-\bA_i)$. This is illustrated in Fig.~\ref{fig.1} for the case $k=2$. In the replica limit $n\to 0$, the off-diagonal matrix 
elements along the first row map onto a piecewise constant function $g(u)$ with $k$ steps on the interval $u\in[0,1)$. Since the level $k$ of RSB is {\it a priori} not known
and could in principle be infinite, we only assume that $g(u)$ is a monotonic function and parametrize the Parisi matrix by $[\tilde{g},g(u)]$. 

We point out that RSB in the quartic terms resembles a weaker form of disorder as compared to conventional spin glasses such as the Edwards-Anderson and Sherrington-Kirkpatrick \cite{Edwards+75,Sherrington+75,Parisi79} or random-field Ising models \cite{Mezard+92}, in which RS is broken on the quadratic level. 
Since  on-site disorder in the BH model leads to random-mass rather than random-field disorder the $U(1)$ symmetry is preserved and RSB can only occur on the 
quartic level, as seen in previous studies of random-mass disorder in ferromagnets \cite{Dotsenko+95a,Dotsenko+95b}. 

\section{Renormalization-group analysis}
We proceed with an RG analysis of the effective replica field theory (\ref{eq.replica_action}) to one-loop order. After successively eliminating modes with momenta from the 
infinitesimal shell $e^{-\ud \mathcal{\ell}}\le |\bk|\le 1$ and rescaling of momenta ($\bk \to \bk e^{\ud\ell}$), frequencies 
($\omega\to\omega e^{z\ud\ell}$, and fields ($\psi_\alpha\to \psi_\alpha e^{-\eta\ud\ell}$) we obtain RG equations for $I_0=1/(1+\overline{R})$ and the rescaled 
Parisi matrix $[\tilde{\lambda},\lambda(u)]=I_0^2 [\tilde{g},g(u)]$, 
\begin{subequations}
\begin{eqnarray}
\frac{\ud I_0}{\ud \ell} & = &  \left[-2+\tilde{\lambda}+\rho \left(\tilde{\lambda}-\langle \lambda \rangle\right)\right]I_{0}+2I_{0}^{2}, \\
\frac{\ud \tilde{\lambda}}{\ud \ell} & = &  (4 I_0 -d)\tilde{\lambda} + 6 \tilde{\lambda}^{2}+2\rho_0 \left[\left(\tilde{\lambda}-\langle \lambda \rangle\right) \tilde{\lambda}\right.\\
& & \left. +2\left(\tilde{\lambda}^2-\langle \lambda^{2} \rangle\right) \right], \nn\\ 
\frac{\ud \lambda(u)}{\ud \ell} & = & (4 I_0 -d) \lambda (u) + 2 \lambda^2(u) +4 \tilde{\lambda}\lambda(u)+4\rho_0\\ 
& & \times \left[\frac52  \left(\tilde{\lambda}-\langle \lambda \rangle\right) \lambda(u)- \int_{0}^{u} \ud v \left[\lambda(u)-\lambda(v)\right]^{2} \right] \nn
\end{eqnarray}
\label{eq.g(u)}
\end{subequations}

Here we have used the multiplication rules \cite{Mezard+91} for Parisi matrices (which form a group) to compute the square 
of $[\tilde{\lambda},\lambda(u)]$. For any $k$-step 
RSB the number of steps and the step positions remain fixed under the RG and the integral-differential equation (\ref{eq.g(u)}c) for $\lambda(u)$ turns into $k+1$ coupled 
ordinary differential equations for the step heights $\lambda_0,\ldots, \lambda_k$. We have defined $\langle \lambda^i \rangle =\int_0^1 \ud u\lambda^i(u)$ and the constant
$\rho_0=\int_\omega [1+I_0 f(\omega)]^{-1}$ which is proportional to the boson filling $m=\int_\omega \mathcal{G}_0(\omega)$. Because of the scaling behavior 
of $I_0$ and $f(\omega)$, $\partial_\ell f(\omega)=(2-z\omega\partial_\omega)f(\omega)$, $\rho_0\sim m$ does not scale under the RG.

\section{Results}
\begin{figure}[t!]
\begin{center}
\includegraphics[width=  \linewidth]{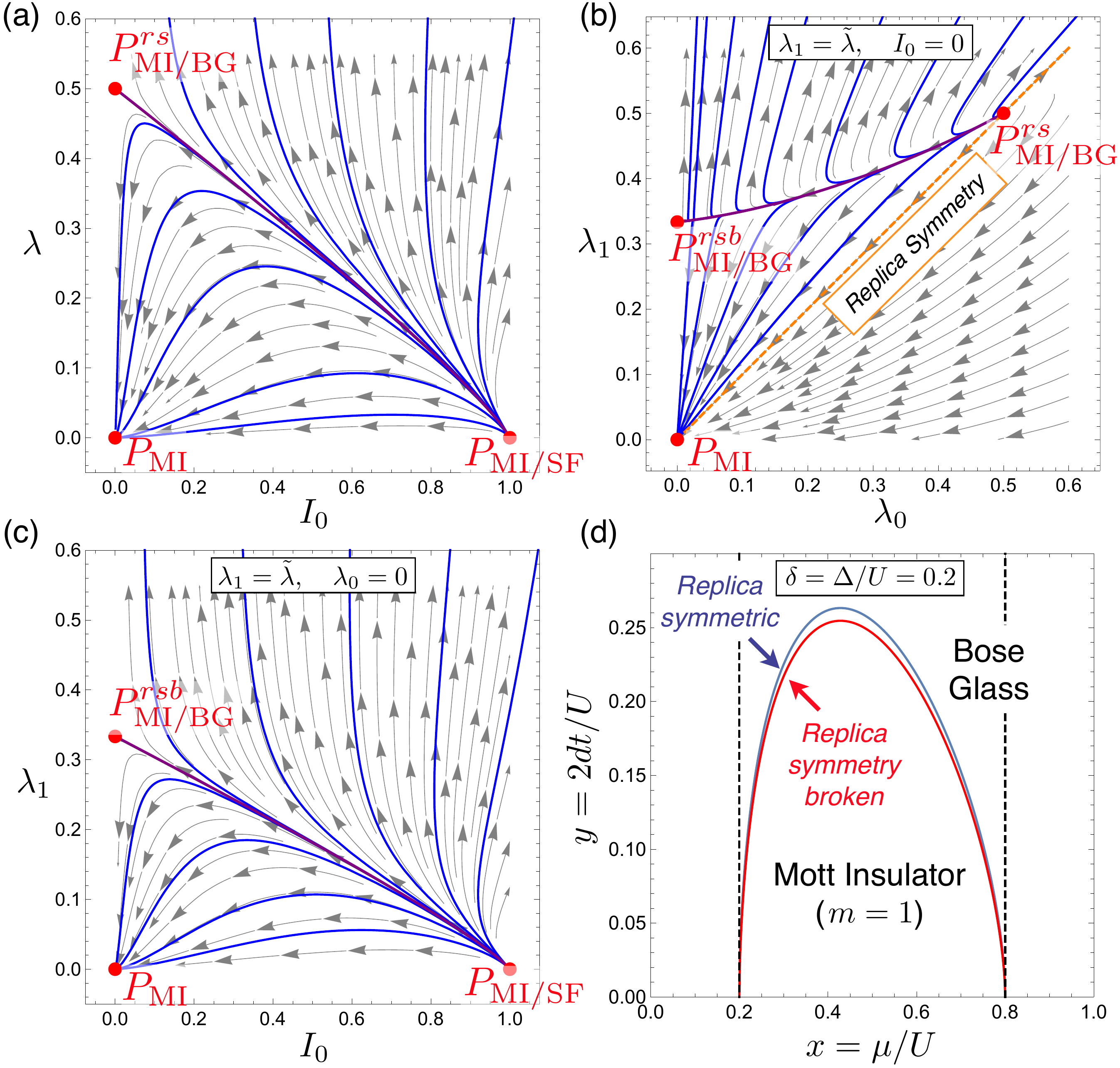}
\caption{(Color online) RG flow in the RS case and $d=3$ as a function of inverse mean $I_0=1/(1+\overline{R})$ and relative variance 
$\lambda=I_0^2 g$ of the random mass distribution. The instability of the MI/SF fixed point ($I_0=0$, $\lambda=0$) against disorder is 
controlled by a separatrix that terminates at the critical fixed point $P_\textrm{MI/BG}$ ($I_0=0$, $\lambda=d/6$) of the MI-to-BG transition. 
 (b) RG flow of the step heights $\lambda_0$ and $\lambda_1$ of the one-step RSB 
Parisi function. The RS fixed point $P^{rs}_\textrm{MI/BG}(d/6,d/6)$ for the MI/BG transition is unstable to RSB with a new fixed point $P^{rsb}_\textrm{MI/BG}(0,d/9)$. 
Note that for $\lambda_0 > \lambda_1$ disorder is irrelevant and the RG flow is towards the MI fixed point. 
(c) RG flow diagram for $\lambda_1$ and $I_0$ for the one-step RSB case with $\lambda_0=0$. (d) Phase boundaries between the first MI lobe and the BG as a function 
of chemical potential $x=\mu/U$ and hopping $y=2dt/U$ for a disorder strength 
of $\delta=\Delta/U=0.2$. The phase boundary for the one-step RSB transition is shown in red, and the boundary for the RS transition in blue.}
\label{fig.2}
\end{center}
\end{figure}

We first revisit the RG flow of the RS system, $\lambda:=\tilde{\lambda}\equiv\lambda(u)$, which has been studied previously \cite{Kruger+11} and is shown 
in Fig.~\ref{fig.2}(a). In this case the RG equations simplify to $I_0'(\ell) = (-2+\lambda)I_{0}+2I_{0}^{2}$ and $\lambda'(\ell) = (4 I_0 -d)\lambda + 6 \lambda^2$. It has 
been pointed out \cite{Kruger+11,Hegg+13} that $\lambda=I_0^2 g$ rather than $g$ is the right variable to distinguish the BG from the MI: in both insulating 
phases the mean of the mass distribution diverges under the RG, $\overline{R}\to\infty$ ($I_0\to 0$), reflecting that SF correlations are short ranged. Since the 
variance $g$ of the mass distribution 
also diverges, we have to compare the shift of the distribution with its spread, which immediately identifies the {\it relative variance} $\lambda$ as the correct variable. 
In the MI, $\lambda\to 0$, indicating that disorder is irrelevant, while in the BG, $\lambda$ diverges.  At a scale $\ell^*$ where $\lambda(\ell^*)\simeq 1$, the left 
tail of the distribution pushes through zero, signaling the presence of rare SF regions in the system. The typical separation of such regions, 
$\xi\simeq e^{\ell^*}$, diverges at the transition to the MI as $\xi\simeq (t-t_c)^{-\nu}$ with an exponent $\nu=1/d$. Note that this scale has nothing to 
do with the SF correlation length.

We now turn to the full set of RG equations (\ref{eq.g(u)}) and search for non-trivial fixed points with RSB. Following previous work \cite{Wu98,DeCesare+99}
we set Eq.~(\ref{eq.g(u)}c) to zero and take a derivative with respect to $u$. Solutions to the resulting equation are given by $\lambda'(u)=0$ or 
\begin{eqnarray}
0 & = & 4I_0-d+4\lambda(u)+4\tilde{\lambda}+10\rho_0 \left(\tilde{\lambda}-\langle\lambda\rangle   \right)\nn\\
& & -8\rho_0\left(u \lambda(u)-\int_0^u\ud v \lambda(v)   \right),
\end{eqnarray}
which after taking another derivative with respect to $u$ yields $\lambda'(u)(1-2\rho_0 u)=0$. This  implies that the Parisi function has to be 
constant for all values of $u$ except for $u_1=1/(2\rho_0)$. Therefore, only RS or one-step RSB fixed points are possible at 
one-loop order. Specializing to a step-function $\lambda(u)=\lambda_0$ for $u\le u_1$ and $\lambda(u)=\lambda_1=\tilde{\lambda}$ for $u>u_1$, 
the RG equations (\ref{eq.g(u)}) reduce to
\begin{subequations}
\label{eq.rg}
\begin{eqnarray}
I_0'(\ell) & = &  (-2+\lambda_1)I_0+\frac12 (\lambda_1-\lambda_0)I_{0}+2I_{0}^{2}, \\
\lambda_0'(\ell) & = & (4I_0-d)\lambda_0-3\lambda_0^2+9\lambda_0\lambda_1,\\
\lambda_1'(\ell) & = & (4I_0-d)\lambda_1-2\lambda_0^2 -\lambda_0\lambda_1+9\lambda_1^2.
\end{eqnarray}
\end{subequations}

It is clear from the RG equations that if the step height is initially zero, $\lambda_0=\lambda_1$, the 
system will remain RS on all scales. We now address the question if an infinitesimal RSB 
perturbation is relevant and if the fixed point $P^{rs}_\textrm{MI/BG}$ becomes unstable towards a new RSB fixed point.

For $I_0=0$, the RG equations ($\ref{eq.rg}$) indeed exhibit a RSB fixed point $P^{rsb}_\textrm{MI/BG}(\lambda_0=0, \lambda_1=d/9)$. 
The RG flow in the $\lambda_0$-$\lambda_1$-plane is controlled by a seperatrix from the RS fixed point 
$P^{rs}_\textrm{MI/BG}(\lambda_0=\lambda_1=d/6)$ to $P^{rsb}_\textrm{MI/BG}$ [see Fig.~\ref{fig.2}(b)]. Below this line and for 
$\lambda_1\ge\lambda_0$ the flow is towards the attractive fixed point $P_\textrm{MI}(\lambda_0=\lambda_1=0)$. In the disordered BG phase infinitesimally 
close to the MI the trajectories follow the seperatrix and the divergence of $\lambda_1$ is controlled by the new RSB fixed point $P^{rsb}_\textrm{MI/BG}$. 
Disorder with a decreasing Parisi step function ($\lambda_0>\lambda_1$) is always irrelevant.

The RG flow in the $I_0$-$\lambda_1$ plane for $\lambda_0=0$ is shown in Fig.~\ref{fig.2}(c). While the stability region of the MI state is reduced compared to 
the RS case [see Fig.~\ref{fig.2}(a)], the overall behavior is very similar.  Linearizing around the critical RSB fixed point  $P^{rsb}_\textrm{MI/BG}$, we obtain
the RG equation $\delta \lambda_1'(\ell)=d\cdot \delta\lambda_1$ for the deviation $\delta \lambda_1:=\lambda_1-d/9$ from the critical point, giving rise to the same correlation length exponent $\nu=1/d$ as in the RS situation.

The phase boundary between the MI and the BG is given by the separatrix connecting $P_\textrm{MI/SF}$ and $P_\textrm{MI/BG}$. Since the initial values of 
$I_0=1/(1+\overline{R})$ and $\lambda=I_0^2 (\overline{R^2}-\overline{R}^2)$ are functions of the chemical potential $x=\mu/U$, the hopping amplitude 
$y=2dt/U$, and the 
disorder strength $\delta=\Delta/U$, it is straightforward to obtain the phase diagram as a function of these parameters. For the box distribution $p(\epsilon_i)$ of 
random on-site energies $\epsilon_i$ the integrals for the mean and variance of the induced random mass distribution $R_i$ (\ref{eq.random_mass}) can be 
calculated analytically. In Fig.~\ref{fig.2}(d) the phase boundaries of the first Mott lobe ($m=1$) for the RS and one-step RSB cases are shown 
as a function of $x$ and $y$ for a disorder strength $\delta=0.2$. While the transition points $x=0.2$ and $x=0.8$ at zero hopping are determined by the 
disorder bound $\delta$, the extent of the Mott lobe depends on the initial momentum cut-off.

\section{Edwards-Anderson order parameters}
Motivated by the need for a measurable signature of RSB, we calculate the Edwards-Anderson order parameters for the BG. Since  different replicas are 
coupled only on the quartic level in the form of a density-density interaction, the glassiness of the disordered BH model is associated with boson number 
fluctuations \cite{Morrison+08}. 
In the following 
we define the density on site $i$ and at time $\tau$, $\rho_i(\tau)=I_{0}^{-1}|\psi_i(\tau)|^2$ where the normalization is such that $\langle \rho_i(\tau) \rangle=\rho_0\sim m$ 
in the MI state. The order parameter signifying RSB \cite{Binder+86} is given by the difference between
\begin{subequations}
\begin{eqnarray}
q_\textrm{EA} & = & \lim_{\tau\to\infty} \left(\overline{\langle \rho_i(\tau)\rho_i(0)  \rangle}- \overline{\langle \rho_i(\tau) \rangle}\phantom{.}\overline{\langle \rho_i(0) \rangle}\right),\\
q & = & \overline{\langle \rho_i(\tau) \rangle^2 }- \overline{\langle \rho_i(\tau) \rangle}^2.
\end{eqnarray}
\label{EA}
\end{subequations}
These functions are related to the connected local density-density correlation function $K_{\alpha\beta}(\tau)=  \llangle \rho_\alpha(\tau)\rho_\beta(0)\rrangle$ in the replica theory.
Note that we have suppressed the site index for brevity. 
To be more specific, 
\begin{equation}
q_\textrm{EA}=\lim_{\tau\to\infty} K_{\alpha\alpha}(\tau)
\end{equation}
measures the long-time correlations within the same replica, whereas 
\begin{equation}
q= \lim_{n\to0}\sum_{\alpha\neq\beta}K_{\alpha\beta}(0)/n(n-1)
\end{equation} 
is determined by the equal-time correlations between different replicas. Note that the relevance of density-density correlations between different replicas and the possibility 
to measure such correlations in optical lattice experiments are discussed in detail in Ref.~\cite{Morrison+08}. At one-loop order we obtain
$q_\textrm{EA}=\rho_0^2\tilde{\lambda}$ and $q= \rho_0^2 \langle \lambda \rangle$. As expected, both order parameters vanish in the MI and are finite in the BG. 
In a RS system, $q=q_\textrm{EA}$, but in the presence of RSB, the two order parameters will become different and $\Delta=q_\textrm{EA}-q$ is related 
to the degree of ergodicity breaking \cite{Binder+86}. For the present one-step RSB solution we obtain $\Delta=\rho_0 (\lambda_1-\lambda_0)/2$.

\section{Susceptibility towards RSB}
The RG flow in the BG is clearly unstable to RSB, $\Delta>0$. However, in this strong-coupling regime the disorder strength increases under the RG,  eventually rendering
the procedure invalid. To provide additional evidence for RSB we therefore calculate the susceptibility associated with an infinitesimal RSB perturbation in the MI. In this
regime the RG is exact since there is an attractive weak-coupling fixed point, guaranteeing that disorder remains small on all length scales. A divergent susceptibility at the critical 
point provides strong evidence  that the strong coupling phase indeed corresponds to a replica symmetry broken solution \cite{LeDoussal+95}. The susceptibility to RSB can be defined
as 
\begin{equation}
\chi_\textrm{RSB} = \lim_{\substack{\ell\to\infty \\ \epsilon(0) \to 0}}  \frac{\Delta(\ell)}{\epsilon(0)},
\end{equation}
where $\epsilon(0)$ is an infinitesimal RSB perturbation such that for $\epsilon(0)\to0$ the system approaches the RS critical point 
$P^{rs}_\textrm{MI/BG}(\lambda_0=\lambda_1=d/6)$ from the MI side, and $\Delta(\ell)$ denotes the RSB order parameter defined above in terms of 
density-density correlation functions. 

\begin{figure}[t!]
\begin{center}
\includegraphics[width=  \linewidth]{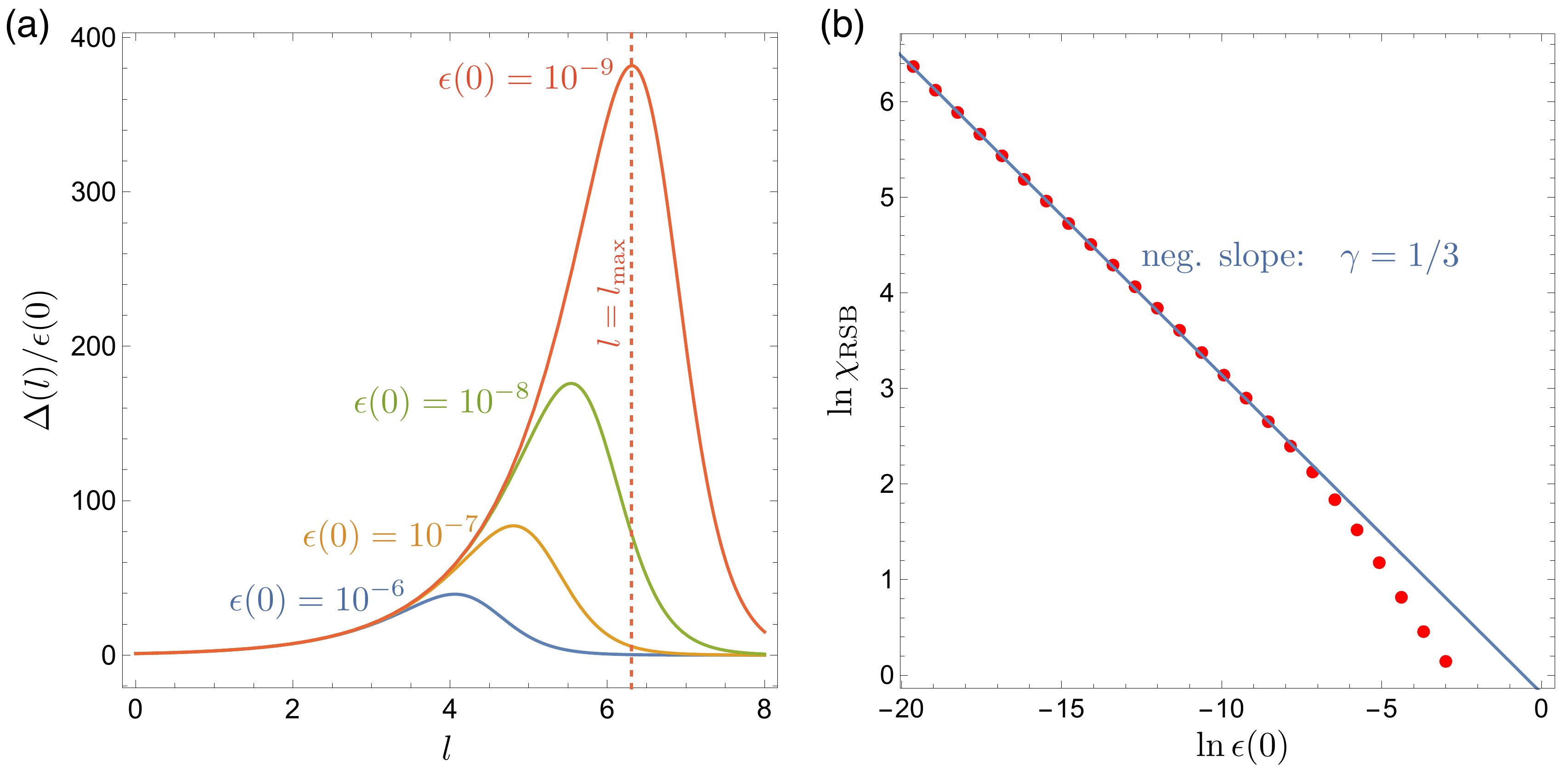}
\caption{(Color online)(a) RG flow in $d=3$ of the RSB order parameter $\Delta(\ell)$ for different infinitesimal RSB perturbations $\epsilon(0)$. As $\epsilon(0)\to0$ the system 
approaches the RS fixed point from the MI phase. (b) Susceptibility $\chi_\textrm{RSB}$ as a function of  $\epsilon(0)$, confirming that  $\chi_\textrm{RSB}\sim \epsilon(0)^{-1/3}$ at the transition.}
\label{fig.3}
\end{center}
\end{figure}

A one-step perturbation that fulfills the above criteria is given by $\lambda_1(0)=\frac{d}{6}-\frac{4}{3}\epsilon(0)$ and $\lambda_0(0)=  \frac{d}{6}-\frac{7}{3}\epsilon(0)$.
In Fig.~\ref{fig.3}(a) the RG flow of $\Delta(\ell)/\epsilon(0)$ is shown for different values of the RSB perturbation $\epsilon(0)$. For any finite $\epsilon(0)>0$,
the RG flow is towards the MI fixed point $(\lambda_1=\lambda_0=0)$ and hence $\Delta(\ell)\to 0$ as $\ell\to \infty$. However, to compute $\chi_\textrm{RSB}$ we simultaneously 
have to take the limit $\epsilon(0)\to0$. A well defined procedure is to evaluate the susceptibility  at the scale $\ell_\textrm{max}$ where the function $\Delta(\ell)/\epsilon(0)$
has its maximum. At this scale, the trajectories escape the close proximity of the seperatrix between the RS and RSB critical points. Both, the position and the hight of this maximum 
diverge as $\epsilon(0)\to 0$. We obtain $\chi_\textrm{RSB}\sim \epsilon(0)^{-\gamma}$ with $\gamma=1/3$ in 
$d=3$ [see Fig.~\ref{fig.3}(b)]. Repeating this calculation for different dimensions $d>2$, we find that $\gamma=1/d$.

While our results unambiguously show that the disordered BH model is inherently unstably to RSB, the question whether infinitesimal RSB perturbations exist in 
the first place remains open. It has been argued by Dotsenko \emph{et al}. \cite{Dotsenko+95b} that small RSB perturbations are always present in disordered systems, including systems 
with random-mass disorder. However, a rigorous proof seems impossible. It is interesting to note that the RG flow of the Cardy-Ostlund, 2$d$ random-field XY model shows 
a similar instability towards RSB \cite{LeDoussal+95} while on the other hand the correlation functions can be computed very accurately by taking a fully RS flow \cite{Perret+12}.

\section{Discussion}
In summary, we have analyzed the transition between the MI and the BG, using an RG analysis of the strong-coupling replica field theory. Our results demonstrate 
that the BG is characterized by RSB which is a manifestation of the lack of ergodicity and self-averaging in the system. We have identified 
Edwards-Anderson type order parameters for the BG, which are associated with glassy fluctuations of the boson density.  

Recent advances in single-atom resolution imaging \cite{Sherson+10,Bakr+10,Weitenberg+11,Endres+11} in cold-atom experiments should enable the detection of the 
local particle-number fluctuations that would distinguish the BG from the MI. Carefully controlled disorder can be generated by, for example, imposing speckle 
potentials \cite{Billy+08,White+09,Pasienski+10,Kondov+11} onto the optical 
lattice or by using a spatial light modulator \cite{Pasienski+08,Bruce+11,Gaunt+12} to vary the lattice depth from site to site. Such experimental control offers the possibility 
to confirm the breakdown of self-averaging in the BG by comparing particle correlations averaged over a large disordered system with the proper disorder 
average over many disorder realizations. In the near future it might also be possible  to create independent copies of the system with the same disorder landscape and 
to measure density-density correlations between these replicas \cite{Morrison+08}. It would also be interesting to investigate density correlations in compressible, Anderson localized fermion phases which have been argued to be rather similar to the BG and shown to exhibit one-step RSB \cite{Giamarchi+01}.

Another promising avenue for measuring the proposed Edwards-Anderson order parameters (\ref{EA}) is given by certain quantum antiferromagnets with 
magnetic easy-plane anisotropy. In such spin systems the bosonic quasiparticles correspond with the magnon excitations and the chemical potential can be tuned by 
a magnetic field perpendicular to the easy plane \cite{Giamarchi+08}. Bose-Einstein condensation of magnons has been predicted theoretically \cite{Giamarchi+99} 
and observed experimentally \cite{Nikuni +00,Ruegg+03}. Very recently, the formation of the analog of the BG phase has been studied in the easy-plane antiferromagnet 
NiCl$_2\cdot4$SC(NH$_2$)$_2$ (DTN) \cite{Yu+12} where disorder has been introduced by chemical substitution. It should be possible in principle to measure 
the BG order parameters through certain spin correlation functions. This would not only be a direct test for RSB but also establish a connection between the BG and 
a wider class of spin-glass phenomena.

\acknowledgments
The authors acknowledge discussions with D. Cassettari, P. Eastham, J. Keeling, C. R\"uegg, and J. Simon. FK acknowledges financial support from EPSRC program TOPNES (EP/I031014/1) 
and EPSRC (EP/I004831/1). SJT acknowledges financial support from the Scottish CM-DTC.


\begin{thebibliography}{10}
\expandafter\ifx\csname url\endcsname\relax\def\url#1{\texttt{#1}}\fi

\bibitem{Fisher+89}
\Name{Fisher M. P.~A., Weichman P.~B., Grinstein G. \and Fisher D.~S.}
  \REVIEW{Phys. Rev. B}{40}{1989}{546}.

\bibitem{Giamarchi+88}
\Name{Giamarchi T. \and Schulz H.~J.} \REVIEW{Phys. Rev. B}{37}{1988}{325}.

\bibitem{Singh+92}
\Name{Singh K.~G. \and Rokhsar D.~S.} \REVIEW{Phys. Rev. B}{46}{1992}{3002}.

\bibitem{Mukhopadhyay+96}
\Name{Mukhopadhyay R. \and Weichman P.~B.} \REVIEW{Phys. Rev.
  Lett.}{76}{1996}{2977}.

\bibitem{Freericks+96}
\Name{Freericks J.~K. \and Monien H.} \REVIEW{Phys. Rev. B}{53}{1996}{2691}.

\bibitem{Svistunov96}
\Name{Svistunov B.~V.} \REVIEW{Phys. Rev. B}{54}{1996}{16131}.

\bibitem{Herbut97}
\Name{Herbut I.~F.} \REVIEW{Phys. Rev. Lett.}{79}{1997}{3502}.

\bibitem{Wu+08}
\Name{Wu J. \and Phillips P.} \REVIEW{Phys. Rev. B}{78}{2008}{014515}.

\bibitem{Kruger+09}
\Name{Kr\"uger F., Wu J. \and Phillips P.} \REVIEW{Phys. Rev.
  B}{80}{2009}{094526}.

\bibitem{Weichman+08}
\Name{Weichman P.~B. \and Mukhopadhyay R.} \REVIEW{Phys. Rev.
  B}{77}{2008}{214516}.

\bibitem{Bissbort+09}
\Name{Bissbort U. \and Hofstetter W.} \REVIEW{Europhys.
  Lett.}{86}{2009}{50007}.

\bibitem{Iyer+12}
\Name{Iyer S., Pekker D. \and Rafael G.} \REVIEW{Phys. Rev.
  B}{85}{2012}{094202}.

\bibitem{Niederle+13}
\Name{Niederle A.~E. \and Rieger H.} \REVIEW{New Journal of
  Physics}{15}{2013}{075029}.

\bibitem{Scalettar+91}
\Name{Scalettar R.~T., Batrouni G.~G. \and Zimanyi G.~T.} \REVIEW{Phys. Rev.
  Lett.}{66}{1991}{3144}.

\bibitem{Krauth+91}
\Name{Krauth W., Trivedi N. \and Ceperley D.} \REVIEW{Phys. Rev.
  Lett.}{67}{1991}{2307}.

\bibitem{Pai+96}
\Name{Pai R.~V., Pandit R., Krishnamurthy H.~R. \and Ramasesha S.}
  \REVIEW{Phys. Rev. Lett.}{76}{1996}{2937}.

\bibitem{Sen+01}
\Name{Sen P., Trivedi N. \and Ceperley D.~M.} \REVIEW{Phys. Rev.
  Lett.}{86}{2001}{4092}.

\bibitem{Lee+01}
\Name{Lee J.-W., Cha M.-C. \and Kim D.} \REVIEW{Phys. Rev.
  Lett.}{87}{2001}{247006}.

\bibitem{Prokof'ev+04}
\Name{Prokof'ev N. \and Svistunov B.} \REVIEW{Phys. Rev.
  Lett.}{92}{2004}{015703}.

\bibitem{Kisker+97}
\Name{Kisker J. \and Rieger H.} \REVIEW{Phys. Rev. B}{55}{1997}{R11981}.

\bibitem{Pollet+09}
\Name{Pollet L., Prokof'ev N.~V., Svistunov B.~V. \and Troyer M.} \REVIEW{Phys.
  Rev. Lett.}{103}{2009}{140402}.

\bibitem{Gurarie+09}
\Name{Gurarie V., Pollet L., Prokof'ev N.~V., Svistunov B.~V. \and Troyer M.}
  \REVIEW{Phys. Rev. B}{80}{2009}{214519}.

\bibitem{Sherson+10}
\Name{Sherson J.~F., Weitenberg C., Endres M., Cheneau M., Bloch I. \and Kuhr
  S.} \REVIEW{Nature}{467}{2010}{68}.

\bibitem{Bakr+10}
\Name{Bakr W.~S., Peng A., Tai M.~E., Ma R., Simon J., Gillen J.~I., F\"olling
  S., Pollet L. \and Greiner M.} \REVIEW{Science}{329}{2010}{547}.

\bibitem{Weitenberg+11}
\Name{Weitenberg C., Endres M., Sherson J.~F., Cheneau M., Schaub P., Fukuhara
  T., Bloch I. \and Kuhr S.} \REVIEW{Nature}{471}{2011}{319}.

\bibitem{Endres+11}
\Name{Endres M., Cheneau M., Fukuhara T., Weitenberg C., Schau\ss P., Gross C.,
  Mazza L., Banuls M.~C., Pollet L., Bloch I. \and Kuhr S.}
  \REVIEW{Science}{334}{2011}{200}.

\bibitem{Billy+08}
\Name{Billy J., Josse V., Zuo Z., Bernard A., Hambrecht B., Lugan P., Cl\'ement
  D., Sanchez-Palencia L., Bouyer P. \and Aspect A.}
  \REVIEW{Nature}{453}{2008}{891}.

\bibitem{White+09}
\Name{White M., Pasienski M., McKay D., Zhou S., Ceperley D. \and DeMarco B.}
  \REVIEW{Phys. Rev. Lett.}{102}{2009}{055301}.

\bibitem{Pasienski+10}
\Name{Pasienski M., McKay D., White M. \and DeMarco B.} \REVIEW{Nat.
  Phys.}{6}{2010}{677}.

\bibitem{Kondov+11}
\Name{Kondov S.~S., McGehee W.~R., Zirbel J.~J. \and DeMarco B.}
  \REVIEW{Science}{334}{2011}{66}.

\bibitem{Pasienski+08}
\Name{Pasienski M. \and DeMarco B.} \REVIEW{Opt. Express}{16}{2008}{2176}.

\bibitem{Bruce+11}
\Name{Bruce G.~D., Mayoh J., Smirne G., Torralbo-Campo L. \and Cassettari D.}
  \REVIEW{Phys. Scr.T}{143}{2011}{014008}.

\bibitem{Gaunt+12}
\Name{Gaunt A.~L. \and Hadzibabic Z.} \REVIEW{Scientific
  Reports}{2}{2012}{721}.

\bibitem{Vojta10}
\Name{Vojta T.} \REVIEW{J. Low Temp. Phys.}{161}{2010}{299}.

\bibitem{Kruger+11}
\Name{Kr\"uger F., Hong S. \and Phillips P.} \REVIEW{Phys. Rev.
  B}{84}{2011}{115118}.

\bibitem{Hegg+13}
\Name{Hegg A., Kr\"uger F. \and Phillips P.} \REVIEW{Phys. Rev.
  B}{88}{2013}{134206}.

\bibitem{Binder+86}
\Name{Binder K. \and Young A.~P.} \REVIEW{Rev. Mod. Phys.}{58}{1986}{801}.

\bibitem{Fischer+91}
\Name{Fischer K.~H. \and Hertz J.~A.} \Book{Spin Glasses} (Cambridge University
  Press) 1991.

\bibitem{Parisi79}
\Name{Parisi G.} \REVIEW{Phys. Rev. Lett.}{43}{1979}{1754}.

\bibitem{Dotsenko+95a}
\Name{Dotsenko V. \and Feldman D.~E.} \REVIEW{J. Phys. A: Math
  Gen}{28}{1995}{5183}.

\bibitem{Dotsenko+95b}
\Name{Dotsenko V., Harris A.~B., Sherrington D. \and Stinchcombe R.~B.}
  \REVIEW{J. Phys. A: Math Gen}{28}{1995}{3093}.

\bibitem{Giamarchi+96}
\Name{Giamarchi T. \and Le~Doussal P.} \REVIEW{Phys. Rev. B}{53}{1996}{15206}.

\bibitem{Giamarchi+01}
\Name{Giamarchi T., Le~Doussal P. \and Orignac E.} \REVIEW{Phys. Rev.
  B}{64}{2001}{245119}.

\bibitem{Sengupta+05}
\Name{Sengupta K. \and Dupuis N.} \REVIEW{Phys. Rev. A}{71}{2005}{033629}.

\bibitem{Edwards+75}
\Name{Edwards S.~F. \and Anderson P.~W.} \REVIEW{J. Phys. F}{5}{1975}{965}.

\bibitem{Sherrington+75}
\Name{Sherrington D. \and Kirkpatrick S.} \REVIEW{Phys. Rev.
  Lett.}{32}{1975}{1792}.

\bibitem{Mezard+92}
\Name{M\'ezard M. \and Young. A.~P.} \REVIEW{Europhys. Lett.}{18}{1992}{653}.

\bibitem{Mezard+91}
\Name{M\'ezard M. \and Parisi G.} \REVIEW{J. Phys. I}{1}{1991}{809}.

\bibitem{Wu98}
\Name{Wu X.} \REVIEW{Physica A}{251}{1998}{309 }.

\bibitem{DeCesare+99}
\Name{De~Cesare L. \and Mercaldo M.~T.} \REVIEW{Phys. Rev. B}{60}{1999}{2976}.

\bibitem{Morrison+08}
\Name{Morrison S., Kantian A., Daley A.~J., Katzgraber H.~G., Lewenstein M.,
  B\"uchler H.~P. \and Zoller P.} \REVIEW{New Journal of
  Physics}{10}{2008}{073032}.

\bibitem{LeDoussal+95}
\Name{LeDoussal P. \and Giamarchi T.} \REVIEW{Phys. Rev. Lett.}{74}{1995}{606}.

\bibitem{Perret+12}
\Name{Perret A., Ristivojevic Z., LeDoussal P., Schehr G. \and Wiese K.~J.}
  \REVIEW{Phys. Rev. Lett.}{109}{2012}{157205}.

\bibitem{Giamarchi+08}
\Name{Giamarchi T., R\"uegg C. \and Tchernyshyov O.} \REVIEW{Nat.
  Phys.}{4}{2008}{198}.

\bibitem{Giamarchi+99}
\Name{Giamarchi T. \and Tsvelik A.~M.} \REVIEW{Phys. Rev. B}{59}{1999}{11398}.

\bibitem{Nikuni+00}
\Name{Nikuni T., Oshikawa M., Oosawa A. \and Tanaka H.} \REVIEW{Phys. Rev.
  Lett.}{84}{2000}{5868}.

\bibitem{Ruegg+03}
\Name{R\"uegg C., Cavadini N., Furrer A., G\"udel H.-U., Kr\"amer K., Mutka H.,
  Wildes A., Habicht K. \and Vorderwisch P.} \REVIEW{Nature}{423}{2003}{62}.

\bibitem{Yu+12}
\Name{Yu R., Yin L., Sullivan N.~S., Xia J.~S., Huan C., Paduan-Filho A.,
  {Oliveira Jr} N.~F., Haas S., Steppke A., Miclea C.~F., Weickert F.,
  Movshovich R., Mun E.-D., Scott B.~L., Zapf V.~S. \and Roscilde T.}
  \REVIEW{Nature}{489}{2012}{379}.

\end{thebibliography}
\end{document}